\label{key}
\documentclass[11pt,a4paper]{article}
\usepackage[latin9]{inputenc}
\usepackage{bm}
\usepackage{amsmath}
\usepackage{amssymb}
\usepackage{graphicx}
\def\ie{\emph{i.e.}, }

\makeatletter

\pdfoutput=1
\usepackage{jcappub}
\usepackage{amsthm}
\usepackage{color}
\usepackage{graphicx}
\usepackage[export]{adjustbox}

\usepackage{array}
\newcolumntype{D}[1]{>{\let\newline\\\arraybackslash\hspace{0pt}}m{#1}} 
\newcolumntype{C}[1]{>{\centering\let\newline\\\arraybackslash\hspace{0pt}}m{#1}}
\newcommand{\ensAve}[1]{\big\langle #1 \big\rangle}

\title{\boldmath A comprehensive study of Modulation effects on CMB Polarization}

\author[a]{Rahul Kothari}
\affiliation[a]{Department of Physics \& Astronomy, University of the 
Western Cape, Cape Town 7535, South Africa}
\emailAdd{quantummechanicskothari@gmail.com}

\abstract{The Cosmic Microwave Background is characterized by temperature and linear polarization fields. Dipole modulation in the temperature field  has been extensively studied in the context of hemispherical power asymmetry. In this article, we show that a dipole modulation, and in general, any kind of modulation isn't allowed in the $E$ and $B$ modes. This is the main result of this paper. This result explains why no evidence of modulation in $E$ mode has been found in the literature. On the contrary, the linear polarization fields $Q$ and $U$ have  no such restrictions. We show that modulation under certain situations can be thought of as local $U(1)$ gauge transformations on the surface of a sphere. As far as the modulation function is concerned, we show that physical considerations enforce it to be (i) a spin $0$ field and (ii) a scalar under parity.  As masking is a specific type of modulation, our study suggests that a direct  masking of $E$ mode isn't also possible. Masking in $E$ map can only be applied through $Q$ and $U$ fields. This means that in principle, leaking of $E$ and $B$ mode powers into each other is unavoidable.
}

\makeatother
\makeatletter
\gdef\@fpheader{}
\makeatother
\begin{document}
\maketitle\flushbottom

\section{Introduction\label{sec:Introduction}}

The Cosmic Microwave Background (CMB hereafter) has been a very important probe to test our cosmological models, thereby improving our understanding of the Universe. CMB comprises of the photons that got decoupled out of the cosmic plasma after the epoch of \emph{recombination} and have been free streaming since then. This epoch corresponds to a redshift of  $z\sim 1000$, \ie when the universe was only 400,000 years old.

This primordial radiation is characterized by temperature field $T$ and linear polarization fields $Q$ and $U$. Although CMB radiation has linear polarization, it doesn't have circular polarization component. This is on account of the fact that circular polarization can't be generated by Thomson scattering \cite{Zaldarriaga:1996xe}. However, some non standard interactions can
generate the same \cite{Sadegh:2017rnr,Vahedi:2018abn,Bartolo:2019eac}. For the purpose of this article, we ignore circular polarization. 

\begin{figure}
\begin{center}
\includegraphics[scale=0.6]{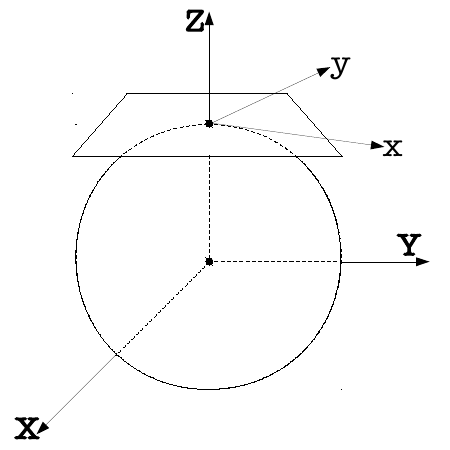}
\par\end{center}
\caption{\label{fig:Rotation-in-the}Rotation in the local coordinate system on the surface of a sphere. Here \emph{XYZ} represents the coordinate axes. Radiation is coming along the $-Z$ axis towards an observer at the origin of the local coordinate system \emph{xyZ}.}
\end{figure}

It is known that $T$ remains invariant  under the rotation of a local coordinate system, i.e., it is a spin 0 field, see Figure \ref{fig:Rotation-in-the}. On the other hand,  
the linear combinations $Q\pm iU$ are respectively spin\footnote{In general, a  field $\Psi$ on a sphere $\mathbb{S}^2$ has spin $s$, if under a right handed rotation of the local coordinate system by an angle $\alpha$, it transforms as $\Psi\mapsto\Psi'=\Psi e^{-is\alpha}$.} $\pm2$ fields.    However, with an appropriate application of the differential operator `eth' ($\eth$) on $Q\pm iU$, we can obtain $E$ and $B$ modes \cite{Zaldarriaga:1996xe}. These modes do behave as scalars under this rotation.

One of the founding pillars of modern cosmology is the \emph{Cosmological Principle}. According to this principle, the Universe is statistically isotropic and homogeneous on length scales $\gtrsim$ Mpc. For the CMB temperature field, this means that the ensemble average of the 2 point correlation function, {viz.}, $\big\langle T(\mathbf{m})T(\mathbf{n}) \big\rangle$ depends only on  the angle between the two directions $\mathbf{m}$ and $\mathbf{n}$, i.e.,  $\big\langle T(\mathbf{m})T(\mathbf{n}) \big\rangle\propto\mathbf{m}\cdot\mathbf{n}$. In other words, the correlation is the same, irrespective of the location of the vectors $\mathbf{m}$ and $\mathbf{n}$ if the angle between them remains unchanged.

Prior to WMAP data release and its subsequent analysis in 2003, it was assumed that CMB radiation satisfies the cosmological principle quite well. The analysis of the WMAP data revealed a power difference in different hemispheres. This implied a potential violation of the cosmological principle at $3\sigma$ level \cite{Eriksen:2003db} and later came to be known as  \emph{Hemispherical Power Asymmetry}.
This effect has still persisted in Planck and WMAP data sets at $\sim 3.0\sigma$ level \cite{Hoftuft:2009rq,Eriksen:2007pc,Ade:2013nlj,Ade:2015hxq}. Planck 2018 results \cite{Planck:2019evm} are consistent with these findings. Phenomenologically, it has been studied using a \emph{dipole
modulation} model \cite{Gordon:2005ai,Gordon:2006ag,Prunet:2004zy,Bennett:2010jb}. 

In the multipole space, the cosmological principle implies that all non-diagonal correlations are zero \cite{Hajian:2003qq}. This is no longer true in the presence of modulation. For example, a dipole modulation leads to non diagonal correlations between $\ell$ and $\ell\pm 1$ multipoles. In the literature, the presence of these non-diagonal correlations has been explained on the basis of modification of the primordial power spectrum \cite{Pullen:2007tu,Kothari:2015tqa,Kothari:2015xva,Ackerman:2007nb,Chang:2018msh,Dey:2013tfa,Ma:2011ii}. 
Similar analyses have also been performed for polarization. Aluri and Shafieloo \cite{Aluri:2017cna} used Planck 2015 polarization maps and found a power asymmetry over the range $\ell =20-240$ when fitted with  a dipole. The direction was found to be broadly aligned with the CMB dipole. Later however, Ghosh and Jain \cite{Ghosh:2018apx} using pixel based method found no evidence of the dipolar modulation signal. In order to test the signal, Planck Team \cite{Planck:2019evm} employed two methods but also didn't find a strong signal of this asymmetry. Kochappan et. al. \cite{Kochappan:2021fza} used contour Minkowski Tensor and Directional Statistic and also found no  statistical deviation from statistical isotropy in the $E$ mode. In this paper, we give a mathematical explanation of this null result. This also happens to be one of  the main results of this paper. We show that $E$ mode, in fact, permits no modulation in general and dipole modulation in particular. Since masking can be considered as a special type of modulation (in a sense to be discussed later), our analysis suggests direct masking in $E$ mode is also not possible. Our finding is consistent with the literature where masking in the $E$ maps has been shown to cause problems \cite{Smith:2005gi,Planck:2019evm}.  Thus the only way of applying masking to $E$ is through $Q$ and $U$ maps. A similar result is true for  $B$ modes as well.

Additionally, it has been well documented that masking of  $Q$ and $U$ leads to a power leakage \cite{Lewis:2001hp, Planck:2019evm}. Thus we also conclude that, in principle, it is not possible to avoid leakage of $E$ and $B$ mode powers upon mask application.

The article is structured in the following manner. We begin in Section
\ref{sec:Preliminaries} discussing important mathematical properties of the harmonic coefficients of CMB temperature and polarization fields. This would be useful in all the subsequent analysis and setting up the notation. This is followed by a precise meaning of modulation and its ramifications for both linear polarization ($Q$ and $U$) and scalar  ($E$ and $B$)  modes in Section \ref{sec:Modulation}. Different aspects like mean, Gaussianity, transformation under parity, etc., of the modulated harmonic coefficients are also discussed. We also discuss some of the desirable properties that the modulation function should satisfy. After this, in Section \ref{sec:Mask_App}, we discuss the consequences of the results obtained in this paper on masking. We conclude in Section  \ref{sec:Conclusion}. Additionally, in Appendix \ref{sec:ModFuncHarCoef}, we provide Table \ref{table:harmonicInfo} of  harmonic coefficients of the modulating function having pure dipolar, quadrupolar, octupolar, etc., modulations. These may find applications elsewhere. \textbf{Notations:} (a) Throughout this article quantities with tilde over them denote modulated quantities (b) unless otherwise stated, all repeated indices are summed over.

\section{Preliminaries\label{sec:Preliminaries}}
Our results in this paper rest on specific properties of polarization fields'  harmonic coefficients. So we first perform a careful analysis of this aspect.  As was discussed before, the temperature field $T$ and the linear combinations $Q\pm iU$ are respectively spin 0 and $\pm2$ fields. The temperature field $T$, thus admits the following spherical harmonic decomposition in terms of the spin 0 spherical harmonics $Y_{\ell m}(\mathbf{n})$.
\begin{equation}
T(\mathbf{n})=\sum_{\ell m}T_{\ell m}Y_{\ell m}(\mathbf{n}),\ \ \sum_{\ell m}\equiv\sum_{\ell=0}^{\infty}\sum_{m=-\ell}^{\ell}\label{eq:T_Harm_Decom}
\end{equation}
On the other hand, the spherical harmonic decomposition of the linear combinations $Q\pm iU$ is expressed as
\begin{equation}
Q(\mathbf{n})\pm iU(\mathbf{n})=\sum_{\ell m}a_{\pm2,\ell m}\ {}_{\pm2}Y_{\ell m}(\mathbf{n}),\label{eq:QIU_Harm_Decom}
\end{equation}
where ${}_{\pm2}Y_{\ell m}$ are spin spherical harmonics \cite{Zaldarriaga:1996xe,Newman1966a,Goldberg1966}. Using the orthogonality properties of spin weighted  harmonics,  this equation can be inverted
\begin{equation}
	a_{\pm2,\ell m}=\int(Q\pm iU)\ {}_{\pm2}Y_{\ell m}^{*}d\Omega.\label{eq:sph_pm2_har}
\end{equation}
Now, the spin harmonics ${}_{s}Y_{\ell m}$ satisfy the property that ${}_{s}Y_{\ell m}=0$ when $\ell<|s|$ \cite{doi:10.1063/1.527183,Dray:1984gy,Newman1966a,Goldberg1966}. This implies that  $a_{\pm2,\ell m}=0$ when $\ell<2$. 

We can also obtain spin 0 fields from the linear combinations $Q\pm iU$ by appropriately applying the  differential operator $\eth$
\begin{equation}
\bar{\eth}^{2}(Q+iU)=\sum_{\ell m}\sqrt{\frac{\left(\ell+2\right)!}{\left(\ell-2\right)!}}a_{2,\ell m}Y_{\ell m},\label{eq:ethsq_q_plus_u}
\end{equation}
\begin{equation}
\eth^{2}(Q-iU)=\sum_{\ell m}\sqrt{\frac{\left(\ell+2\right)!}{\left(\ell-2\right)!}}a_{-2,\ell m}Y_{\ell m}.\label{eq:ethsq_q_minus_u}
\end{equation}
Notice that in Eqs. \eqref{eq:ethsq_q_plus_u} and \eqref{eq:ethsq_q_minus_u}, although the sum doesn't contribute for $\ell<2$ (due to negative factorial in the denominator), this doesn't guarantee that $a_{\pm2,\ell m}=0$. Instead, the conclusion is reached on the basis of properties of spin spherical harmonics ${}_{\pm 2}Y_{\ell m}$. The scalar $E$ and $B$ modes are defined as linear combinations \cite{Zaldarriaga:1996xe} of Eqs. \eqref{eq:ethsq_q_plus_u} and \eqref{eq:ethsq_q_minus_u}
\begin{equation}
E=-\frac{1}{2}\left[\bar{\eth}^{2}(Q+iU)+\eth^{2}(Q-iU)\right].\label{eq:E_Mode_Defi}
\end{equation}
\begin{equation}
B=\frac{i}{2}\left[\bar{\eth}^{2}(Q+iU)-\eth^{2}(Q-iU)\right].\label{eq:B_Mode_Defi}
\end{equation}
Furthermore, $E$ being a scalar, admits the following harmonic decomposition \cite{Zaldarriaga:1996xe}
\begin{equation}
E=\sum_{\ell m}\sqrt{\frac{\left(\ell+2\right)!}{\left(\ell-2\right)!}}E_{\ell m}Y_{\ell m}.\label{eq:E_Harm_Decom}
\end{equation}
At this point, we would like to point out the difference between harmonic expansions of $T$ and $E$ fields given respectively in Eqs. \eqref{eq:T_Harm_Decom} and \eqref{eq:E_Harm_Decom}. The harmonic coefficients $T_{\ell m}\ne 0$ in general for any given $\ell$. On the other hand, $E$ mode harmonic coefficients $E_{\ell m}$ are necessarily 0 when $\ell<2$. This can be seen by relating $E_{\ell m}$'s with $a_{\pm2,\ell m}$ 
\begin{equation}
E_{\ell m}=-\frac{1}{2}\left(a_{2,\ell m}+a_{-2,\ell m}\right).\label{eq:emode_harm_coeff}
\end{equation}

A similar conclusion holds for $B$ mode harmonic coefficient
\begin{equation}
B_{\ell m}=\frac{i}{2}\left(a_{2,\ell m}-a_{-2,\ell m}\right).\label{eq:bmode_harm_coeff}
\end{equation}
From this analysis, we conclude that for any physical $E$ ($B$) field, we must have $E_{\ell m}=0$ ($B_{\ell m}=0$) when $\ell <2$. This property is true by definition.

\section{Modulation Analyses  \label{sec:Modulation}}
The power asymmetry in CMB temperature field has been studied using the following dipole modulation \cite{Gordon:2005ai,Gordon:2006ag,Prunet:2004zy,Bennett:2010jb} model 
\begin{equation}
	\tilde{T}(\mathbf{n})=T(\mathbf{n})(1+A\bm{\lambda}\cdot\mathbf{n}).\label{eq:t_dip_mod}
\end{equation}
In Eq. \eqref{eq:t_dip_mod}, the modulated field $\tilde{T}$ is obtained from the statistically isotopic temperature field $T$ after multiplying it by a dipole modulation term. This modulation is characterized by the magnitude $A$ and a preferred direction $\bm{\lambda}$ that violates statistical isotropy.  This  motivates the following definition. For a given field $\Xi$ (real or complex), a \emph{modulation} is the transformation $\Xi(\mathbf{n})\mapsto\tilde{\Xi}(\mathbf{n})=f(\mathbf{n})\Xi(\mathbf{n})$, where $f(\mathbf{n})$ is a specific \emph{modulating function} (real or complex). 

In the purview of this definition, masking can be thought of as a special case of modulation. Masking is applied to a map in order to remove regions which aren't a part of a survey or when these have to be deleted because of various contaminations. Mask value is usually taken to be zero for a region intended to be removed. But modulation allows any general angle $\mathbf{n}\equiv (\theta,\phi)$ dependent function $f(\mathbf{n})$ subjected to the following two physical requirements
\begin{itemize}
\item It shouldn't change the spin of the field $\Xi$. This implies that the modulation function $f$ can only have spin 0. This is a consequence of the fact that the total spin of the product of two fields with spins $s_1$ and $s_2$ is  $s_1+s_2$.
	
\item It should transform as a scalar under parity, i.e., $\mathbf{n}\mapsto -\mathbf{n}$. The dipole modulation in Eq. \eqref{eq:t_dip_mod} contains two vectors (polar vectors to be precise) $\mathbf{n}$  and a fixed direction $ \bm{\lambda} $. Using a dot product of two polar vectors, we can only construct a scalar. Since dipole modulation is special case of $f(\mathbf{n})$, so we demand that it  also transforms as a scalar under parity.
\end{itemize}

In the remainder of this section, we first study the modulation in the linear polarization fields $Q$ and $U$ and then in the scalar modes $E$ and $B$. 

\subsection{Linear Polarization Q and U}\label{sec:QU_Modu} 
It is useful to study modulation directly in spin $\pm2$ fields $Q\pm iU$. This, as we will see, includes a `direct' modulation in $Q$ and $U$ fields as a special case. Thus we write the modulation as (a similar equation exists for $\tilde{Q}-i\tilde{U}$ but it won't give us any new information)
\begin{equation}
	\tilde{Q}+i\tilde{U} = (Q+iU)f. \label{eq:QPlusiUMod}
\end{equation}
To keep the analysis as general as possible, we  take  complex modulating function $f=f_1+if_2$, with $f_1$ and $f_2$ being real. This can also be written in the form
\begin{equation}
	\tilde{Q}+i\tilde{U} = (Q+iU)Re^{i\Psi}\label{eq:RPsi}
\end{equation}
From Eq. \eqref{eq:RPsi},  it is clear that a modulation transforms the complex field $Q+iU\mapsto \tilde{Q}+i\tilde{U}$ with the simultaneous application of (a) scaling by $R=\sqrt{f_1^2+f_2^2}$ and (b) an anticlockwise rotation in the tangent plane by $\Psi=\tan^{-1}(f_2/f_1)$  at each point $\mathbf{n}\in \mathbb{S}^2$. It is interesting to note that a similar rotation ensues in $Q\pm iU$ on account of addition of Chern Simons term in the electromagnetic Lagrangian \cite{Greco:2022ufo}. To gain more insights, Eq. \eqref{eq:QPlusiUMod} can be written in the following matrix form
\begin{equation}
	\begin{pmatrix}\tilde{Q}\\
		\tilde{U}
	\end{pmatrix}=\begin{pmatrix}f_1 & -f_2\\
		f_2 & f_1
	\end{pmatrix}\begin{pmatrix}Q\\
		U
	\end{pmatrix}\label{eq:mat_repre}
\end{equation}
We can now consider two special cases
\begin{itemize}
\item Consider a specific transformation for which the scaling $|f|=R=1$. Then the modulation corresponds to a local $U(1)$  gauge transformation of $Q+iU$ on $\mathbb{S}^2$.  

\item Now consider the case when modulating field is real, \ie $f_2=0$. Using Eq. \eqref{eq:mat_repre}, we get $\tilde{Q}=f_1 Q$ and $\tilde{U}=f_1 U$ which represents a modulation in the individual fields. Thus the modulation suggested in Eq. \eqref{eq:QPlusiUMod} is more general and  includes  direct modulation in $Q$ and $U$ as a special case. 
\end{itemize}

Now we study the consequences of the modulation in Eq. \eqref{eq:QPlusiUMod}. We notice that our formalism  generalizes the dipole modulation studies \cite{Ghosh:2015qta,Contreras:2017zjv,Ghosh:2018apx} performed in the context of CMB polarization. For this, it is useful to express $Q\pm iU$ in terms of $E$ and $B$ mode harmonic coefficients.
This can be done using Eqs. \eqref{eq:emode_harm_coeff} \& \eqref{eq:bmode_harm_coeff}
and we get
\begin{equation}
	Q\pm iU=-\sum_{\ell m}(E_{\ell m}\pm iB_{\ell m})\ _{\pm2}Y_{\ell m}\label{eq:Q-U-func-E-B}
\end{equation}
We notice that the modulated fields $\tilde{Q}\pm i\tilde{U}$ are also spin $\pm2$. This is true  since $f$ must have spin 0 (see last paragraph of Section \ref{sec:Modulation}). Therefore when we multiply it with $Q\pm iU$ (spin $\pm 2$ fields), we again get spin $\pm 2$. So we can again expand $\tilde{Q}\pm i\tilde{U}$ in terms of spin $\pm2$ spherical harmonics as per Eq. \eqref{eq:QIU_Harm_Decom}. Repeating the same steps as above, the corresponding equation for the modulated fields would be
\begin{equation}
	\tilde{Q}\pm i\tilde{U}=-\sum_{\ell m}(\tilde{E}_{\ell m}\pm i\tilde{B}_{\ell m})\ _{\pm2}Y_{\ell m}.\label{eq:Mod-Q-U-func-Mod-E-B}
\end{equation}

After some simplifications, the  modulated $E$ and $B$ mode harmonic coefficients, corresponding to the modulation \eqref{eq:QPlusiUMod}, are expressed as
\begin{equation}
\tilde{E}_{\ell m}=\frac{(-1)^{m}}{2}{\sum_{\ell_im_i}} f_{\ell_{2}m_{2}}\mathcal{G}_{\ell,\ell_{1},\ell_{2};-2,2,0}^{-m,m_{1},m_{2}}\left[\left(E_{\ell_{1}m_{1}}+iB_{\ell_{1}m_{1}}\right)+(-1)^{L}\left(E_{\ell_{1}m_{1}}-iB_{\ell_{1}m_{1}}\right)\right]\label{eq:Elm_QU_mod}
\end{equation}
\begin{equation}
	\tilde{B}_{\ell m}=\frac{(-1)^{m}}{2i}{\sum_{\ell_im_i}} f_{\ell_{2}m_{2}}\mathcal{G}_{\ell,\ell_{1},\ell_{2};-2,2,0}^{-m,m_{1},m_{2}}\left[\left(E_{\ell_{1}m_{1}}+iB_{\ell_{1}m_{1}}\right)-(-1)^{L}\left(E_{\ell_{1}m_{1}}-iB_{\ell_{1}m_{1}}\right)\right]\label{eq:Blm_QU_mod}
\end{equation}
In both these equations $L=\ell+\ell_1+\ell_2$ and $\mathcal{G}_{\ell_{1},\ell_{2},\ell_3;s_1,s_2,s_3}^{m_{1},m_{2},m_3}$ is the generalized Gaunt symbol, defined as the integral over three spin spherical harmonics
\begin{eqnarray}
	\mathcal{G}^{m_1,m_2,m_3}_{\ell_1,\ell_2,\ell_3;s_1,s_2,s_3} &=& \int\ _{s_{1}}Y_{\ell_{1}m_{1}}\ _{s_{2}}Y_{\ell_{2}m_{2}}\ _{s_{3}}Y_{\ell_{3}m_{3}}d\Omega \notag\\
	&=& \sqrt{\frac{\prod_{i=1}^{3}(2\ell_{i}+1)}{4\pi}}\begin{pmatrix}\ell_{1} & \ell_{2} & \ell_{3}\\
		m_{1} & m_{2} & m_{3}
	\end{pmatrix}\begin{pmatrix}\ell_{1} & \ell_{2} & \ell_{3}\\
		-s_{1} & -s_{2} & -s_{3}
	\end{pmatrix}
\end{eqnarray}
We notice the presence of the Wigner 3j symbol
\begin{equation}
	\begin{pmatrix}\ell & \ell_{1} & \ell_{2}\\
		2 & {-2} & {0}
	\end{pmatrix}=0
\end{equation}
in Eqs.  \eqref{eq:Elm_QU_mod} and \eqref{eq:Blm_QU_mod}. This symbol guarantees that both $\tilde{E}_{\ell m}$ and $\tilde{B}_{\ell m}$ are zero when $\ell <2$. An alternative way of concluding this is to express the modulated harmonic coefficients using tensor spherical harmonics \cite{Smith:2005gi}. This means that any given  modulation $f$
in $Q$ and $U$ fields gives rise to physically acceptable $E$ and
$B$ fields. These results match with the ones existing in the literature  \cite{Lewis:2001hp,Hansen:2002iha}. We must also point out that Eqs.  \eqref{eq:Elm_QU_mod} and \eqref{eq:Blm_QU_mod} can also be derived using the derivative properties of $\eth$ on spin spherical harmonics in \eqref{eq:ethsq_q_plus_u} and \eqref{eq:ethsq_q_minus_u}. 

We also notice that  the modulation \eqref{eq:QPlusiUMod} intermixes $E$ and $B$ modes. This intermixing is due to modulation \eqref{eq:QPlusiUMod} and is different from the one that arises on account of gravitational lensing \cite{Zaldarriaga:1998ar}. It is known that the $T$ and $E$ modes are sourced by both scalar and tensor perturbations whereas the $B$ modes are generated solely due to tensor perturbations. Thus to detect the primordial $B$ mode generated by tensor perturbations, in addition to removing the effects due to lensing \cite{Kesden:2002ku,Knox:2002pe}, one in principle, must also remove the effects due to modulation.

\subsubsection{Harmonic Coefficients' Behaviour \label{sec:HarmCoeffBehave}}
The spherical harmonic coefficients $X_{\ell m}$ with $X\in\{T,E,B\}$ satisfy various properties. For example, it is known that $X_{\ell m}$, for a given multipole $\ell$ with $m>0$ are distributed as multivariate Gaussian \cite{Upham:2019ruv} with mean $0$, i.e., $\ensAve{X_{\ell m}}=0$. From Eqs. \eqref{eq:Elm_QU_mod} and \eqref{eq:Blm_QU_mod}, it is clear that the modulated coefficients still have zero mean, i.e., $\ensAve{\tilde{E}_{\ell m}}=0$ and $\ensAve{\tilde{B}_{\ell m}}=0$.  Further mod coefficients are also Gaussian distributed \cite{Upham:2019ruv}. Thus modulation doesn't change these statistical properties of $X_{\ell m}$.

We now discuss the properties of these coefficients under parity. Under parity transformation, $T$ and $E$ behave as scalar but $B$ on the other hand, behaves as a pseudo scalar \cite{Planck:2019evm,LiteBIRD:2022cnt,Lewis:2001hp,Lin:2004xy}. Further, it was discussed in Section \ref{sec:Modulation} that the modulation function $f$ can only be a scalar. Based on these facts,  we can conclude that under parity
\begin{eqnarray}
	X_{\ell m}&\mapsto & (-1)^{\ell}X_{\ell m},\ X\in\{T,f,E\}\label{eq:XParity}\\
	B_{\ell m}&\mapsto & -(-1)^\ell B_{\ell m}\label{eq:BParity}
\end{eqnarray}
Now using Eqs. \eqref{eq:XParity} and \eqref{eq:BParity} in Eq. \eqref{eq:Elm_QU_mod}, the modulated $E$ mode harmonic coefficient, under parity, transforms in the following manner
\begin{equation*}
	\tilde{E}_{\ell m}\mapsto  \frac{(-1)^{m}}{2}{\sum_{\ell_im_i}}(-1)^{L+\ell_1+\ell_2} f_{\ell_{2}m_{2}}\mathcal{G}_{\ell,\ell_{1},\ell_{2};-2,2,0}^{-m,m_{1},m_{2}}\left[\left(E_{\ell_{1}m_{1}}+iB_{\ell_{1}m_{1}}\right)+(-1)^{L}\left(E_{\ell_{1}m_{1}}-iB_{\ell_{1}m_{1}}\right)\right]
\end{equation*}
But $L=\ell+\ell_1+\ell_2$, so we get $\tilde{E}_{\ell m} \mapsto(-1)^\ell \tilde{E}_{\ell m}$. This is true for  $B$ mode as well. Thus we conclude that under parity, modulated $E$ and $B$ fields transform in the same manner as unmodulated fields.


\subsubsection{Modulated Power Spectrum}
Cosmological Principle  imposes the following conditions on 2 point correlations in the multipole space
\begin{equation}
	\ensAve{E_{\ell m}E^*_{\ell'm'}}=\delta_{\ell\ell'}\delta_{mm'}C_\ell^{EE},\  \ensAve{B_{\ell m}B^*_{\ell'm'}}=\delta_{\ell\ell'}\delta_{mm'}C_\ell^{BB},\ \ensAve{E_{\ell m}B^*_{\ell'm'}}=0 \label{eq:PowSpecStaIso}
\end{equation}
i.e., only diagonal correlations are non-zero and the last condition follows if the ensemble average is assumed  parity symmetric \cite{Lewis:2001hp}. 

However, in the presence of modulation, we get non-diagonal correlations as well. The general correlation between the modulated harmonic coefficients can be obtained by using statistical isotropy conditions  \eqref{eq:PowSpecStaIso} in Eq. \eqref{eq:Elm_QU_mod}
\begin{eqnarray}
	\ensAve{\tilde{E}_{\ell m}\tilde{E}_{\ell'm'}^*}&=&\frac{(-1)^{m+m'}}{4}\sum_{\ell_im_i}f^*_{\ell_4m_4}f_{\ell_2m_2}\mathcal{G}^{-m,m_1,m_2}_{\ell,\ell_1,\ell_2;-2,2,0}\mathcal{G}^{-m',m_1,m_2}_{\ell,\ell_1,\ell_2;-2,2,0}\notag\\
	& \times & \left[C_{\ell_1}^{EE}(1+(-1)^L)(1+(-1)^{L'})+C_{\ell_1}^{BB}(1-(-1)^L)(1-(-1)^{L'})\right]
\end{eqnarray}
where $L=\ell+\ell_1+\ell_2$ and $L'=\ell'+\ell_1+\ell_4$. It can be seen that the correlation isn't zero even when $\ell\ne\ell'$. Physically, this means that the presence of modulation violates the cosmological principle.  For the special case when $\ell=\ell'$ and $m=m'$, we get 
\begin{eqnarray}
	\ensAve{\tilde{C}^{EE}_\ell}&=&\frac{2\ell+1}{8\pi}\sum_{\ell_2,\ell_2}f_{\ell_2}(2\ell_1+1)(2\ell_2+1)\begin{pmatrix}\ell & \ell_{1} & \ell_{2}\\
		2 & {-2} & {0}
	\end{pmatrix}^2 \notag\\
	&\times & \left[C_{\ell_1}^{EE}(1+(-1)^{\ell+\ell_1,\ell_2})+C_{\ell_1}^{BB}(1-(-1)^{\ell+\ell_1,\ell_2})\right]
\end{eqnarray}
where $f_\ell=\sum_{|m|\le \ell}|f_{\ell m}|^2/(2\ell+1)$. 
This matches with Ref. \cite{Rocha:2009mb,Planck:2019evm}. A similar analysis can be performed for the $ B $ mode as well. We also notice that the cross correlation $\ensAve{\tilde{E}_{\ell m}\tilde{B}^*_{\ell'm'}}$ isn't zero in general and becomes zero only when we take $\ell=\ell'$ and $m=m'$.

\subsection{Scalar Modes E and B\label{sec:ScalarModu}}
Now we study the effects of modulation in scalar modes. The most general modulation in the $E$ mode is of the following type
\begin{equation}
	\tilde{E}(\mathbf{n})=E(\mathbf{n})f(\mathbf{n}).\label{eq:Poss-Mod-E}
\end{equation}
Since $E$  is real field, here we take $f$ to be real as well. All three fields in Eq. \eqref{eq:Poss-Mod-E} are spin 0 and can be expanded in terms of the usual spherical harmonics $Y_{\ell m}$. Using orthogonality and product properties of $Y_{\ell m}$, we can relate the modulated $\tilde{E}_{\ell m}$ coefficients with the unmodulated ones ${E}_{\ell m}$
\begin{equation}
	\tilde{E}_{\ell m}=(-1)^{m}{\sum_{\ell_i m_i}} f_{\ell_{1}m_{1}}E_{\ell_{2}m_{2}}\sqrt{\frac{(2\ell+1)(2\ell_{1}+1)(2\ell_{2}+1)}{4\pi}}\begin{pmatrix}\ell & \ell_{1} & \ell_{2}\\
		0 & 0 & 0
	\end{pmatrix}\begin{pmatrix}\ell & \ell_{1} & \ell_{2}\\
		-m & m_{1} & m_{2}
	\end{pmatrix}.\label{eq:ScalarEModu}
\end{equation}
This equation expresses the modulated harmonic coefficients $\tilde{E}_{\ell m}$ as a linear combination of the unmodulated ones ${E}_{\ell m}$, weighted appropriately by  modulating function harmonic coefficients $f_{\ell m}$. By assumption, the unmodulated $E$ field by itself is physical which means
that $E_{\ell m}=0$ when $\ell<2$ (see Section \ref{sec:Preliminaries}). 

Now to show that the modulation in Eq. \eqref{eq:Poss-Mod-E} is not possible, we calculate a specific harmonic coefficient $\tilde{E}_{10}$ of the modulated field, which after some simplifications gives
\begin{equation}
	\tilde{E}_{10}=\sqrt{\frac{3}{4\pi}}\sum_{\ell m}f_{\ell m}(-1)^{\ell}\left[E_{\ell-1,m}^{*}\sqrt{\frac{\ell^{2}-m^{2}}{4\ell^{2}-1}}-E_{\ell+1,m}^{*}\sqrt{\frac{(\ell+1)^{2}-m^{2}}{(2\ell+1)(2\ell+3)}}\right].\label{eq:E_10_mod_coeff}
\end{equation}
It can be seen that since the harmonic coefficient of the modulating field $f_{\ell m}\ne0$ in general,  $\tilde{E}_{10}$ can't be zero either. This implies that no modulation of the form \eqref{eq:Poss-Mod-E}
is allowed as it leads to mathematical inconsistencies. This is the main result of this paper. 
In particular, dipole modulation isn't allowed. To see this explicitly, we take  $f(\mathbf{n})=1+A\bm{\lambda}\cdot\mathbf{n}$, with $A$ (real) and $\bm{\lambda}\equiv(\Theta,\Phi)$ being respectively the magnitude and direction of the modulation. Using Table \ref{table:harmonicInfo},  the harmonic coefficients are
\begin{equation}
	f_{\ell m}=\sqrt{4\pi}\delta_{\ell0}\delta_{m0}+\delta_{\ell1}\frac{4\pi}{3}AY_{1m}^{*}(\bm{\lambda}).\label{eq:dip_mod_coeff}
\end{equation}
Using \eqref{eq:dip_mod_coeff} in \eqref{eq:E_10_mod_coeff} and
after some simplifications we get
\begin{equation}
	\tilde{E}_{10}=A\sqrt{\frac{2}{5}}\left[E_{21}\sin\Theta\cos\Phi+E_{20}\sqrt{\frac{2}{3}}\cos\Theta \right].
\end{equation}
Since in general $E_{21}\ne 0$ and $E_{20} \ne 0$, $\tilde{E}_{10}\ne 0$ as well. This contradicts the basic property of the $E$  field that $E_{\ell m}=0$ when $\ell <2$. Thus we conclude that a dipole modulation in $E$ mode isn't allowed. 
This explains the null results pertaining to the modulation of the $E$  mode  \cite{Planck:2019evm,Ghosh:2018apx,Kochappan:2021fza}. A similar analysis would hold for the $B$ mode polarization. 


\subsubsection{Properties of Modulation Coefficients}
Although the impossibility of $E$ modulation (Eq. \ref{eq:Poss-Mod-E}) renders the modulated coefficients (Eq. \ref{eq:ScalarEModu}) unphysical, it is still worthwhile to compare the properties of these harmonic coefficients with those obtained in Section \ref{sec:HarmCoeffBehave}. We find that just like Eq. \eqref{eq:Elm_QU_mod}, the modulated coefficients in Eq. \eqref{eq:ScalarEModu} are
\begin{itemize}
	\item  Again Gaussian distributed with zero mean
	\item Expressed as a linear combinations of $E_{\ell m}$ and thus have a similar transformation as that of unmodulated ones under parity
\end{itemize}
From this we can conclude that merely the presence of these properties isn't enough to conclude that Eq. \eqref{eq:Poss-Mod-E} is unphysical. Thus the unphysical nature of this modulation is an independent conclusion. 

\section{Applications to Masking Procedure \label{sec:Mask_App}}
In this section, we apply the hitherto obtained results to the masking procedure. As was discussed before, masking can be considered as a special type of modulation. Thus we conclude that
\begin{itemize}
	\item Masking can't be directly applied to $E$ or $B$  as it will lead to mathematical inconsistencies (see Section \ref{sec:ScalarModu}). This is an auxiliary conclusion of the paper. Although Ref. \cite{Smith:2005gi} suggests an alternative,  it  leads to problems like enhancement of noise power \cite{Planck:2019evm}.
	\item As the masking of $ E $ and $ B $ maps can only be performed through $ Q $ and $ U $ maps, it will, in principle, lead to intermixing of $ E $ and $ B $ mode powers. 
\end{itemize}



\section{Conclusions and Outlook \label{sec:Conclusion}}
In this article, we show that no modulation in $E$ mode is possible. This is on account of the properties of the harmonic coefficients.  This explains the null result related to  $E$ mode dipole modulation. No such restrictions are however for $Q$ and $U$ maps. We find that when the magnitude of the modulating function is unity,  the modulation can be thought of as a local $U(1)$ gauge transformation on $\mathbb{S}^2$. We  have also studied various properties of the modulated harmonic coefficients. We find that the aforementioned conclusion, i.e., the unphysical nature of $E$ mode modulation, can't be reached just on the basis of the statistical properties of harmonic coefficients and is thus an independent conclusion.

Since masking is a special type of modulation, we also conclude that masking in $E$ maps should only be introduced through $Q$ and $U$ maps. This masking, as the literature sufficiently attests, leads to an intermixing $E$ and $B$ mode powers. Thus we find that masking will inevitably lead to intermixing of powers.

We have also found that the modulation function can't be arbitrary. Physics restricts it to have spin 0 and scalar under parity transformation. 

Our results imply that the presence of non-diagonal correlations in the polarization field, suggested in the literature \cite{Kothari:2015tqa}, cannot be attributed to a modulation in the scalar modes $E$ and $B$. However our analysis doesn't rule out the possibility that such correlations can arise due to modified power spectra based models \cite{Pullen:2007tu,Kothari:2015tqa,Kothari:2015xva,Ackerman:2007nb,Chang:2018msh,Dey:2013tfa,Ma:2011ii}  that may further arise due to reasons like spacetime non-commutativity \cite{Akofor:2007fv,Kothari:2015xva}, direction dependent primordial perturbations \cite{Namjoo:2014pqa,Zarei:2014raa}, etc. It would thus be interesting to study the direct connections of modified power spectra on field modulation.  Any kind of inflationary model implying a modulation in $E$ and $B$ is ruled out from the start.

\section*{Data Availability}
No new data was generated or analysed in support of this research.

\section*{Acknowledgements}
 I am thankful to Shamik Ghosh, Prof. Pankaj Jain for illuminating discussions that culminated in  this paper. I am extremely grateful to Prof. Roy Maartens for suggestions. Finally, I am enormously indebted to the anonymous referee whose comments were very helpful in improving the presentation of this paper. I'm supported by the South African Radio Astronomy Observatory (SARAO) and the National Research Foundation (Grant No. 75415). I also sincerely
acknowledge the Institute Post Doctoral Fellowship of IIT Madras where some part of this work was done.

\appendix

\section{Spherical Harmonic Coefficients of the Modulating Function\label{sec:ModFuncHarCoef}}
\begin{figure}
\begin{center}
\includegraphics[scale=0.45]{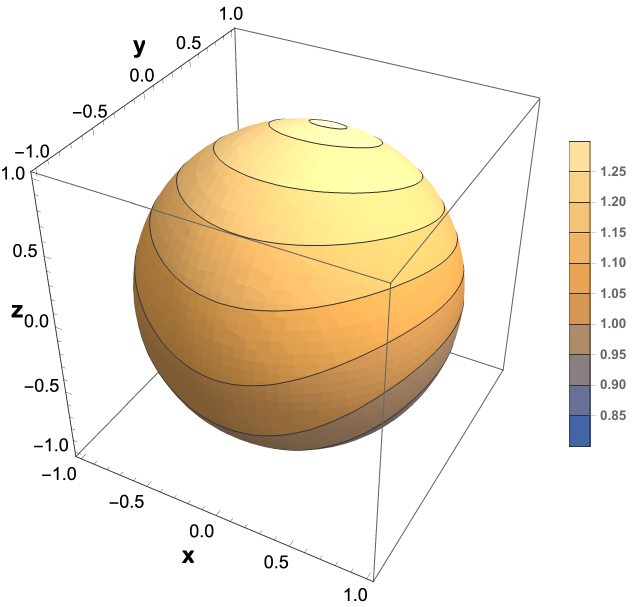}\includegraphics[scale=0.45]{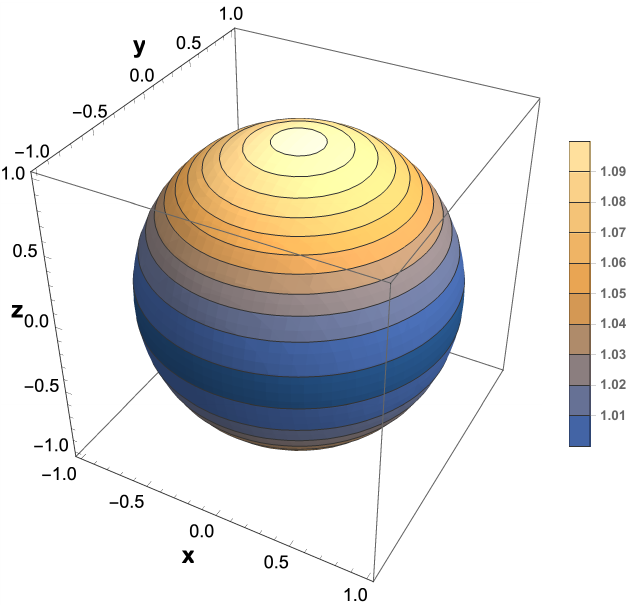}\includegraphics[scale=0.45]{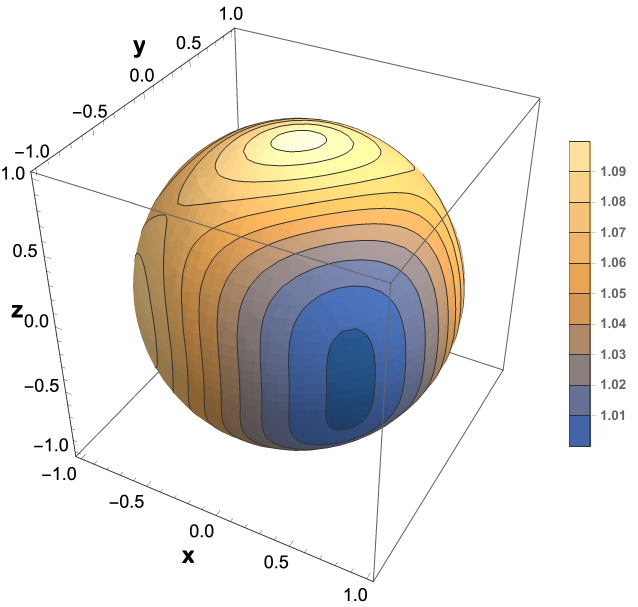}
\par\end{center}
\caption{\label{fig:Modu-Func}Plot of the modulating function $f$ in different
cases. For all these cases, we take $A_0=0$. From left to right,  (a) \emph{a linear combination of dipole and quadrupole} $\to$ $A_{i}=0.2\delta_{1i}+0.1\delta_{2i}$,
$\bm{\lambda}_{i}=(0,0,1)\delta_{1i}+(0,1,1)\delta_{2i}$ (b) \emph{a pure quadrupole} $\to$ $A_{i}=0.2\delta_{2i}$,
$\bm{\lambda}_{i}=(0,0,1)\delta_{2i}$ and (c) \emph{a linear combination of quadrupole
and hexadecapole} $\to$ $A_{i}=0.09\delta_{2i}+0.1\delta_{4i}$,
$\bm{\lambda}_{i}=(-1,-1,0)\delta_{2i}+(0,0,1)\delta_{4i}$.}

\end{figure}
Our analysis till this point  restricts the function $f$ to only have spin 0 and being scalar  under parity. But in principle it can take any form. In this section, we study specific forms of the modulating function $f$.  Our choice is motivated by the dipole modulation model that has been employed to study hemispherical power asymmetry in the $T$ field of CMB (Eq. \ref{eq:t_dip_mod}). A similar kind of dipole modulation has been used for $Q\pm iU$ fields \cite{Ghosh:2015qta,Contreras:2017zjv,Ghosh:2018apx,Namjoo:2014pqa}. This modulation has only one amplitude $A$ and a direction $\bm{\lambda}$. 

In general, we can have different alignments of dipolar, quadrupolar, octupolar, etc., modulations along different directions $\bm{\lambda}_i$ and with different amplitudes $A_i$. These  would be proportional to   different exponents of $\bm{\lambda}_i\cdot \mathbf{n}$. This motivates the following modulating function, 
\begin{equation}
f(\mathbf{n})=1+A_{1}(\bm{\lambda}_1\cdot \mathbf{n})+A_{2}(\bm{\lambda}_2\cdot \mathbf{n})^{2}+\ldots=\sum_{i=0}^{\infty}A_{i}(\cos\gamma_{i})^{i},\ \ A_0=1,\ \ A_i\in \mathbb{C}. \label{eq:specical-case}
\end{equation}
In the above equation, we have defined $\cos\gamma_i=\bm{{\lambda}}_i\cdot\mathbf{n}$. Notice that we have written the modulating function as a linear combination of pure dipole, quadrupole, etc., terms. In Figure \ref{fig:Modu-Func}, we have shown the plots of the modulating function $f$ with various possibilities. 

In order to calculate the corresponding modulated coefficients, our objective is to find out the spherical harmonic coefficients $f_{\ell m}$ of the modulating function $f$. For that, we notice that any power of $\cos\gamma_{i}$ can
be written as a linear combination of the Legendre's polynomials $\mathcal{P}_{\ell}(\cos\gamma_{i})$
with appropriate coefficients. So we write (no sum over $i$ on either
sides)
\begin{equation}
(\cos\gamma_{i})^{i}=\sum_{\ell\ge0}\alpha_{i,\ell}\mathcal{P}_{\ell}(\cos\gamma_{i}).\label{eq:poly_leg_pol_expan}
\end{equation}
The `base change' coefficients $\alpha_{i,\ell}$ can be easily found using any table on Legendre's polynomials. Using addition theorem of spherical harmonics,
we can express the Legendre's polynomials in terms of
spherical harmonics
\begin{equation}
\mathcal{P}_{\ell}(\cos\gamma_{i})=\frac{4\pi}{2\ell+1}\sum_{m=-\ell}^{\ell}Y_{\ell m}(\mathbf{n})Y_{\ell m}^{*}(\bm{\lambda}_{i}).\label{eq:add_thm_sph_harm}
\end{equation}
Finally, using Eqs. \eqref{eq:add_thm_sph_harm} and \eqref{eq:poly_leg_pol_expan} in \eqref{eq:specical-case}, the spherical harmonic coefficients $f_{\ell m}$ are found to be
\begin{equation}
    f_{\ell m} = \frac{4\pi}{2\ell+1}\sum_{i=0}^\infty A_i\,\alpha_{i,\ell}\,Y_{\ell m}^*(\bm{\lambda}_i)\label{eq:f_EllM}
\end{equation}
These harmonic coefficients for some special cases of pure monopole, dipole, etc., modulations are given in Table \ref{table:harmonicInfo}.

\begin{table}
\centering
\begin{tabular}{D{1cm} D{3cm} D{2cm} D{5.5cm}}
\hline
\hline
\noalign{\vskip 0.1cm}
$i$ & Modulation & $\alpha_{i,l}$ & $f_{\ell m}$ \\
\noalign{\vskip 0.1cm}
\hline
\hline
\noalign{\vskip 0.1cm}
$i=0$ & Pure Monopole & $\delta_{0\ell}$ & $\sqrt{4\pi}\delta_{\ell 0}\delta_{m0}$ \\
\hline
\noalign{\vskip 0.1cm}
$i=1$ & Pure Dipole & $\delta_{1\ell}$ & $A_1\dfrac{4\pi}{3}\delta_{\ell1}Y_{1m}^{*}(\bm{\lambda}_1)$ \\
\noalign{\vskip 0.1cm}
\hline
\noalign{\vskip 0.1cm}
$i=2$ & Pure Quadrupole & $\dfrac{\delta_{0\ell}+2\delta_{2\ell}}{3}$ & $A_2\dfrac{4\pi}{3}[\delta_{0\ell}Y^*_{00}(\bm{\lambda}_2)+\frac{2}{5}\delta_{2\ell}Y^*_{2m}(\bm{\lambda}_2)]$ \\
\noalign{\vskip 0.1cm}
\hline
\noalign{\vskip 0.1cm}
$i=3$ & Pure Octopole & $\dfrac{3\delta_{1\ell}+2\delta_{3\ell}}{5}$ & $A_3\dfrac{4\pi}{5}[\delta_{1\ell}Y^*_{1m}(\bm{\lambda}_3)+\frac{2}{7}\delta_{3\ell}Y^*_{3m}(\bm{\lambda}_3)]$ \\
\noalign{\vskip 0.1cm}
\hline
\hline
\end{tabular}
\caption{Spherical harmonic coefficients $f_{\ell m}$ of the modulating function $f$ in some specific cases. In the table, we have also shown the base change coefficients $\alpha_{i,\ell}$.}
\label{table:harmonicInfo}
\end{table}

\bibliographystyle{JHEP}
\bibliography{bibliography}

\end{document}